\documentclass[aps,pra,superscriptaddress,twocolumn,showpacs]{revtex4}
\usepackage{graphicx}
\usepackage{amssymb}
\usepackage{amsmath}
\usepackage{bm}

\usepackage{tikz}
\usetikzlibrary{shapes, positioning}

\usepackage{color}

\begin{document}

\title[]{Complex Wigner entropy and Fisher control of negativity in an oval quantum billiard}

\author{Kyu-Won \surname{Park}}
\email{parkkw7777@gmail.com}
\affiliation{Department of Mathematics and Research Institute for Basic Sciences, Kyung Hee University, Seoul, 02447, Korea}
\author{Jongin \surname{Jeong}}
\email{jijeong2004@gmail.com}
\affiliation{Department of Physics, Pusan National University, Busan, 46241, Korea}
\affiliation{Team QST, Seoul National University, Seoul, 08826, Korea}

\date{\today}

\begin{abstract}
We develop a complex-entropy framework for Wigner negativity and apply it to avoided crossings in an oval quantum billiard. For a real Wigner function the Gibbs--Shannon functional becomes complex; its imaginary part, proportional to the Wigner-negative volume, serves as an entropy-like measure of phase-space nonclassicality. A sign-resolved decomposition separates the total negative weight from its phase-space distribution and defines a negative-channel Fisher information that quantifies how sensitively the negative lobe reshapes as a control parameter is varied. This structure yields a Cauchy--Schwarz bound that limits how rapidly the imaginary entropy, and hence the Wigner negativity, can change with the parameter. In the oval billiard, avoided crossings display enhanced negativity and an amplified negative-channel Fisher response, providing a clear phase-space signature of mode hybridization. The construction is generic and extends to other wave-chaotic and mesoscopic systems with phase-space representations.
\end{abstract}

\maketitle
\section{Introduction}
The Wigner function provides a phase-space representation of quantum and wave states that is formally analogous to a classical phase-space density while retaining full coherence information \cite{Wigner1932,Moyal1949,Hillery1984}. In particular, its negative values are widely interpreted as a hallmark of nonclassicality or non-Gaussian structure \cite{Kenfack2004}. In continuous-variable and optical quantum information, Wigner-negativity based measures have been developed as resource quantifiers for nonclassical light and non-Gaussian states \cite{NonclassicalityReview,Albarelli2018}, and the negativity of generalized Wigner functions has been used as an operational signature in entanglement witnessing and related nonclassicality tests \cite{Arkhipov2018,Chabaud2021}. In wave and quantum billiards, the interlacing of positive and negative lobes can encode detailed information about mode structure that is largely invisible in non-negative intensity distribution alone. \cite{Berry1977,Stockmann1999,Jeong2018PRE}.
 
 Experimentally, recent advances in state tomography and control have made Wigner functions and their negative regions directly accessible in several platforms. In optical continuous-variable systems, homodyne tomography and related techniques enable direct reconstruction of phase-space quasidistributions \cite{Lvovsky2009,Weedbrook2012}, and Wigner-negative features have been observed in a variety of nonclassical light fields \cite{Harder2016,Liu2022,Xiang2022}. In atomic platforms, direct measurements of the Wigner function have been reported for trapped atoms \cite{Winkelmann2022}, while in superconducting circuits propagating Wigner-negative states and cavity-based nonclassicality have been demonstrated \cite{Lu2021,Mohamed2021}. These developments further motivate phase-space based analyses of how eigenmodes reorganize under changes of geometry or boundary conditions.

On the theoretical side, information-theoretic quantities such as Shannon entropies and Fisher information provide a natural language for such analyses~\cite{Shannon1948, CoverThomas}. Entropic measures have been employed to characterize localization, complexity, and parameter sensitivity in atomic, mesoscopic, and wave-chaotic systems \cite{Yukalov2003,DehesaReview,Luzanov2008,He2015}, and they have been applied to avoided crossings in closed and open billiards using configuration- and phase-space probability distributions. \cite{Jeong2018PRE,Park2022Chaos}.  However, most existing work is restricted to nonnegative probability densities such as mode intensity or Husimi functions\cite{Husimi1940, Park2025PRE, Park2022Chaos}. When carried over to Wigner representations this restriction is problematic, because any entropy built from a positive smoothed version of $W$ necessarily discards the sign structure that makes the Wigner function genuinely phase sensitive. This motivates an information-theoretic framework that treats Wigner negativity and its phase-space rearrangements on the same footing.

To pursue this route, we extend the Gibbs--Shannon functional to the complex plane and apply it directly to real Wigner functions, so that its imaginary part is proportional to the Kenfack--\.Zyczkowski negativity and serves as an complex-valued entropy-like measure of nonclassicality \cite{CerfComplexWignerEntropy, Kua2025}. In this work we develop a channel-resolved version of this construction that combines the complex Wigner entropy with Fisher-information measures defined on the positive and negative sign channels. We then apply this framework to an oval quantum billiard with a single shape parameter and use it to analyze how Wigner negativity and its phase-space distribution respond to avoided crossings in the eigenvalue spectrum.

The remainder of the paper is organized as follows. In Sec.~\ref{sec:system} we introduce the oval quantum billiard, describe the numerical eigenmode calculations, and identify the avoided crossing that serves as our testbed. In Sec.~\ref{sec:complex-entropy} we define the complex Wigner entropy, construct the sign-resolved channels, and derive the Fisher control on the negativity dynamics. We then present numerical results for the interacting modes and discuss how the complex entropy and channel Fisher information behave across the avoided-crossing region. Finally, we summarize our findings and outline possible extensions to non-Hermitian and driven systems in the Conclusion.

\section{Oval quantum billiard and avoided crossings}
\label{sec:system}

Quantum billiards provide a  wave model of classically integrable or nonintegrable cavities, in which a free particle or scalar wave is confined to a bounded planar domain \(\Omega\) and obeys the Helmholtz equation with hard-wall boundary conditions,
\begin{equation}
(\nabla^2 + k^2)\,\psi(\mathbf r) = 0,
\qquad
\psi(\mathbf r)\big|_{\mathbf r\in\partial\Omega}=0,
\label{eq:helmholtz}
\end{equation}
which yields a discrete spectrum of real wavenumbers \(k_n\) and eigenmodes \(\psi_n(\mathbf r)\). In the classical limit these systems exhibit mixed or chaotic ray dynamics, while in the wave description they generate dense families of near-degenerate levels and frequent avoided crossings as the shape is varied~\cite{Berry1977,Stockmann1999, Jeong2018PRE}.

In this work the domain \(\Omega\) is an oval billiard obtained by deforming an ellipse in the \(x\) direction. For deformation parameter \(\vartheta=0\) the boundary is an ellipse with semi-axes \(a\) and \(b\); for \(\vartheta\neq 0\) the boundary is taken as
\begin{equation}
\frac{x^2}{a^2}
+\bigl(1+\vartheta x\bigr)\frac{y^2}{b^2}
=1,
\label{eq:oval-boundary}
\end{equation}
so that increasing \(\vartheta\) deforms the ellipse along the \(x\) direction while keeping the overall size comparable~\cite{Park2025PRE}.
For each value of \(\vartheta\), we solve Eq.~\eqref{eq:helmholtz} numerically with Dirichlet boundary conditions on the boundary defined by Eq.~\eqref{eq:oval-boundary}, obtaining the eigenvalues \(k(\vartheta)\) and the corresponding eigenmodes \(\psi(x,y;\vartheta)\) using the boundary element method~\cite{Wiersig2003}.

Among the resulting eigenvalue trajectories we select a near-degenerate pair that undergoes a clear avoided crossing as \(\vartheta\) is varied. The upper and lower branches in Fig.~1 are denoted mode~2 and mode~1, respectively, and the sampling points \(A\!-\!F\) along these curves mark representative values of \(\vartheta\) on both sides of, and at, the avoided crossing. The corresponding intensity patterns \(|\psi(x,y)|^2\) in the lower panels of Fig.~1 show that the two modes exchange their spatial structure across the avoided crossing, providing the configuration-space reference for the phase-space analysis that follows.

\begin{figure}
\centering
\includegraphics[width=8.5cm]{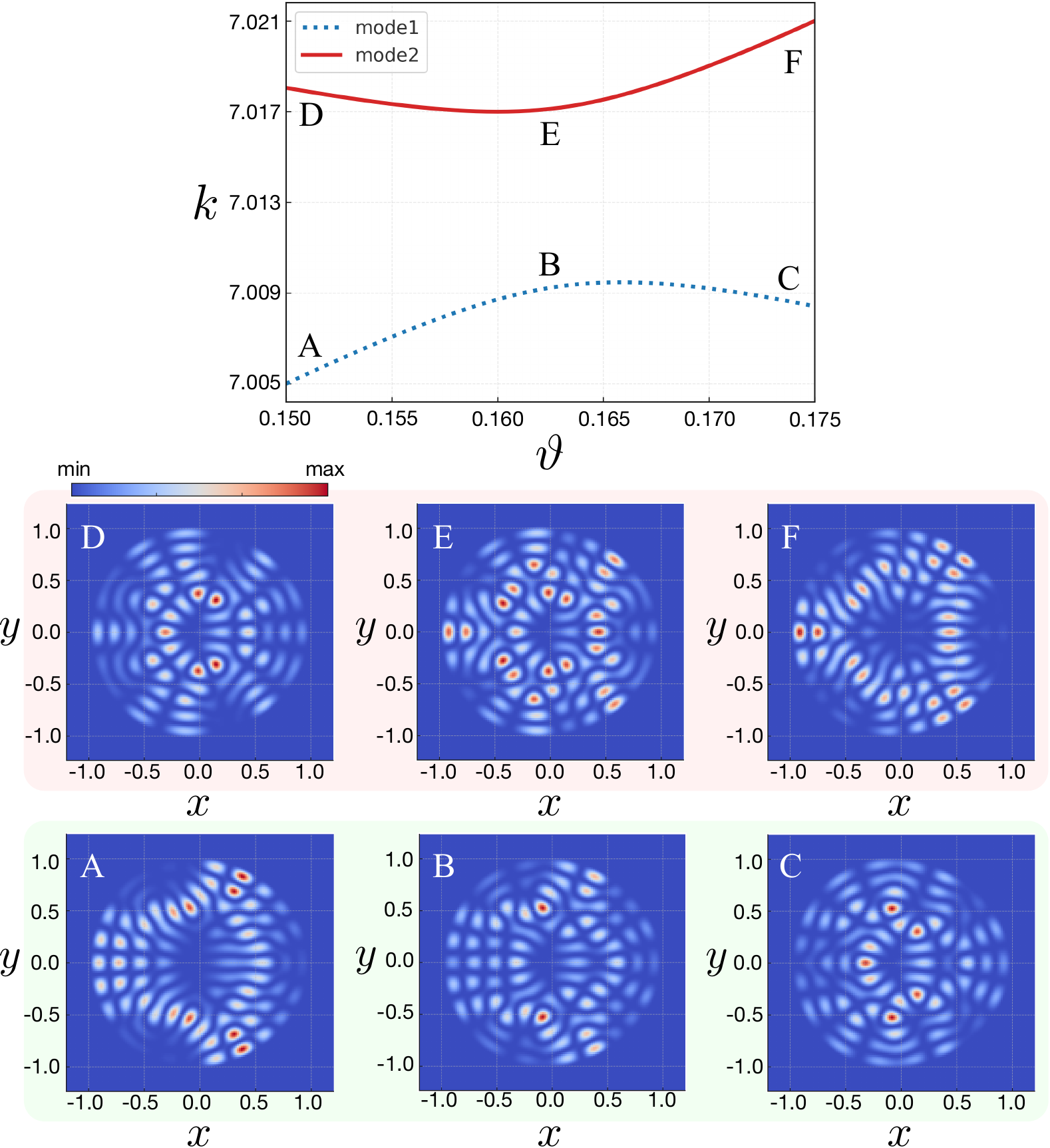}
\caption{
(Top) Eigenvalue trajectories of two interacting modes near the avoided crossing, with sampling points \(A\!-\!C\) (mode\,1) and \(D\!-\!F\) (mode\,2). 
(Bottom) Corresponding intensity patterns \(|\psi(x,y)|^{2}\), showing the exchange of spatial structure between the two modes across the avoided-crossing region.
}
\label{Figure-1}
\end{figure}

\begin{figure*}[htbp]
\centering
\includegraphics[width=17.5cm]{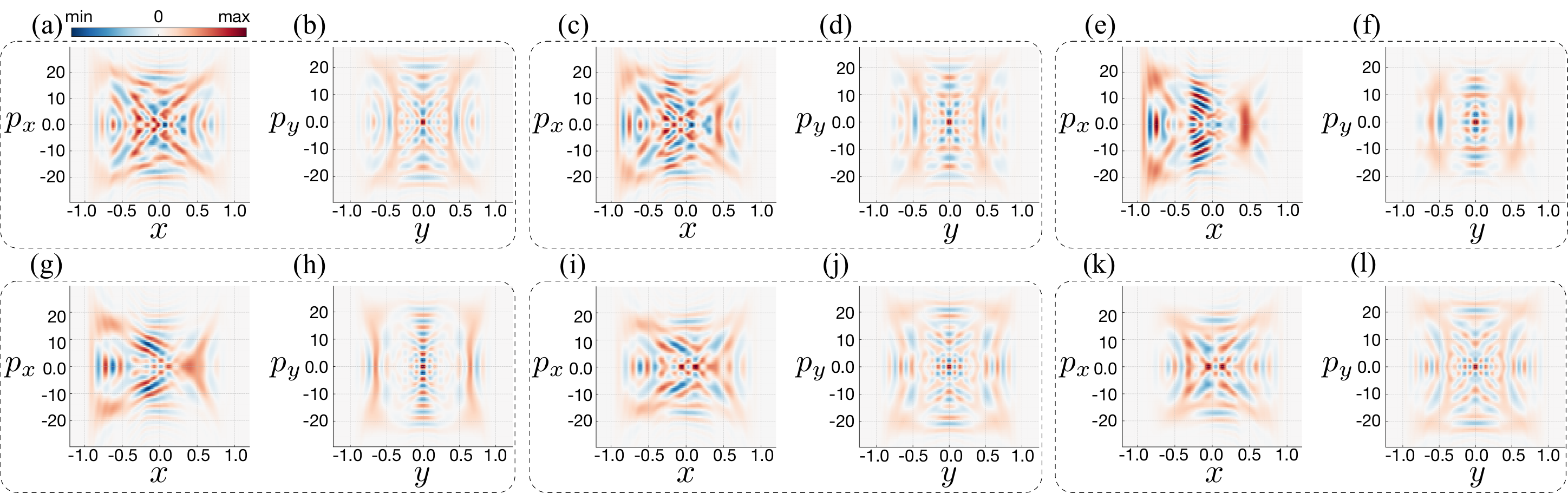}
\caption{
Representative Wigner sections corresponding to the eigenvalue trajectories in Fig.~1. 
Panels (a),(b) show mode\,2 at \(D\); (c),(d) at \(E\); and (e),(f) at \(F\). 
Panels (g),(h) show mode\,1 at \(A\); (i),(j) at \(B\); and (k),(l) at \(C\). 
Each pair displays the phase-space slices \(W_\vartheta(x,y{=}0,p_x,p_y{=}0)\) (left) and \(W_\vartheta(x{=}0,y,p_x{=}0,p_y)\) (right). 
The colormap is centered at zero (red: positive, blue: negative), so blue regions directly visualize Wigner negativity. 
The negative domains intensify and become more finely structured as the system approaches the avoided-crossing region, providing a clear phase-space signature of mode hybridization.
}
\label{Figure-2}
\end{figure*}

\section{Complex Wigner entropy and Fisher control of negativity}
\label{sec:complex-entropy}

The Wigner function provides a phase-space representation of a quantum (or wave) state that is formally analogous to a classical phase-space density yet retains full coherence information. Throughout this section we restrict to pure states with density operator \(\rho_\vartheta = |\psi_\vartheta\rangle\langle\psi_\vartheta|\), where \(\psi_\vartheta(r) = \langle r|\psi_\vartheta\rangle\) denotes the position-space eigenmode at control parameter \(\vartheta\). For such a state we denote by \(W_\vartheta(r,p)\) the real Wigner distribution on phase space \((r,p)\), defined by
\begin{align}
W_\vartheta(r,p)
= \frac{1}{(2\pi\hbar)^2}\int d^2 s\;
e^{\frac{i}{\hbar}p\cdot s}\,
\psi_\vartheta\bigl(r-\tfrac{s}{2}\bigr)\,
\psi_\vartheta^*\bigl(r+\tfrac{s}{2}\bigr),
\end{align}
and normalized by \(\iint W_\vartheta(r,p)\,d^2r\,d^2p = 1\)~\cite{Wigner1932}. The marginals of \(W_\vartheta\) reproduce the configuration- and momentum-space intensities, but unlike a genuine probability density it can take negative values. These negative regions encode interference between different components of the state and are widely interpreted as a measure of nonclassicality or non-Gaussian structure~\cite{Kenfack2004}. 

Information-theoretic diagnostics such as Shannon entropy and Fisher information have become standard tools for quantifying localization, complexity, and parameter sensitivity in quantum and wave systems. In their conventional form they are applied to honest probability densities \(p(z)\ge0\) that are normalized, \(\int p(z)\,dz = 1\). For such positive \(p\) the continuous Shannon entropy is defined by
\begin{align}
H[p] = -\int p(z)\,\ln p(z)\,dz,
\end{align}
while the corresponding Fisher information with respect to a control parameter \(\vartheta\) is
\begin{align}
\mathcal{F}[p_\vartheta] 
= \int p_\vartheta(z)\,\bigl[\partial_\vartheta \ln p_\vartheta(z)\bigr]^2 dz,
\end{align}
which measures how sensitively the distribution \(p_\vartheta\) responds to changes of \(\vartheta\) and sets the scale of optimal parameter estimation through the Cramér--Rao bound~\cite{Shannon1948, CoverThomas,Paris2009,Alipour2014}.

For phase-space representations this positivity restriction is problematic, any entropy built directly from \(|\psi_\vartheta|^2\) or from a Husimi function discards the sign structure that makes \(W_\vartheta\) genuinely phase sensitive. To retain the full information content of \(W_\vartheta\), including its negativity, one is naturally led to extend Gibbs--Shannon functionals to the complex plane~\cite{CerfComplexWignerEntropy, Kua2025}. In this section we construct such a complex Wigner entropy, separate its real and imaginary parts into sign-resolved channels, and show how a Fisher information defined on the negative support controls the parameter dependence of the Wigner negativity, especially near avoided crossings.

Before turning to the formal definitions, it is useful to visualize the phase-space structure of the avoided crossing. Figure~\ref{Figure-2} displays representative Wigner sections associated with the six sampling points \(A\!-\!F\) on the eigenvalue trajectories in Fig.~1. The upper row shows the Wigner function of mode~2 at points \(D\), \(E\), and \(F\), while the lower row shows mode~1 at \(A\), \(B\), and \(C\). Each pair of panels corresponds to two orthogonal cuts through phase space, \(W_\vartheta(x,y{=}0,p_x,p_y{=}0)\) on the left and \(W_\vartheta(x{=}0,y,p_x{=}0,p_y)\) on the right, which sample the phase-space structure along the coordinate axes. The colormap is centered at zero so that red regions indicate positive Wigner weight and blue regions indicate negative weight. As the system approaches the avoided-crossing center, the blue domains become more pronounced and more finely structured, already signaling that Wigner negativity is amplified and strongly rearranged by hybridization. The complex-entropy and Fisher analysis below provides a quantitative framework for this behavior.

\subsection{Complex Wigner entropy and channel structure}

We start from the Gibbs--Shannon functional applied directly to the real Wigner function \(W_\vartheta(r,p)\),
\begin{equation}
\mathcal{H}[W_\vartheta]
= -\iint d^2r\,d^2p\;
W_\vartheta(r,p)\,\ln W_\vartheta(r,p),
\label{eq:complex-entropy-def}
\end{equation}
and write \(\mathcal{H}[W_\vartheta]=h_r(\vartheta)+i\,h_i(\vartheta)\).
The complex logarithm is taken on the principal branch~\cite{CerfComplexWignerEntropy, Kua2025}. Writing
\(\ln W_\vartheta=\ln|W_\vartheta|+i\arg W_\vartheta\) with
\(\arg W_\vartheta\in(-\pi,\pi]\), and using the fact that \(W_\vartheta\) is real, we have
\(\arg W_\vartheta=0\) on the positive support and \(\arg W_\vartheta=\pi\) on the negative support. Substituting into Eq.~\eqref{eq:complex-entropy-def} gives
\begin{align}
h_r(\vartheta)
&= -\iint d^2r\,d^2p\;
W_\vartheta(r,p)\,\ln|W_\vartheta(r,p)|,
\label{eq:hr-def}
\\[2pt]
h_i(\vartheta)
&= \pi\iint_{W_\vartheta<0}\! d^2r\,d^2p\;
|W_\vartheta(r,p)|.
\label{eq:hi-def}
\end{align}
It is therefore natural to define the negative volume
\begin{equation}
N(\vartheta)
= \iint_{W_\vartheta<0}\! d^2r\,d^2p\;
|W_\vartheta(r,p)|,
\label{eq:neg-def}
\end{equation}
for which \(h_i(\vartheta)=\pi\,N(\vartheta)\).
This negative volume is directly related to the Kenfack--\.Zyczkowski Wigner negativity,
and serves as a standard nonclassicality monotone.
The imaginary entropy $h_i(\vartheta)$ is simply a rescaled version of the negativity $N(\vartheta)$ and isolates the contribution of genuinely nonclassical interference.  While $h_i(\vartheta)$ provides the overall magnitude of nonclassicality, to fully understand the local phase-space dynamics and geometric structure of nonclassical interference, it is necessary to examine its spatial extent.

To resolve how this negativity is distributed in phase space, we decompose the Wigner function into sign-resolved channels,
\(
W_{\vartheta,+}(r,p)=\max\{W_\vartheta(r,p),0\}
\)
and
\(
W_{\vartheta,-}(r,p)=\max\{-W_\vartheta(r,p),0\},
\)
with channel masses
\(
Z_\pm(\vartheta)=\iint W_{\vartheta,\pm}(r,p)\,d^2r\,d^2p,
\)
which satisfy
\(Z_+(\vartheta)-Z_-(\vartheta)=1\) and \(Z_-(\vartheta)=N(\vartheta)\).
We factor out these global weights and introduce the shape probabilities
\(P_{\vartheta,\pm}(r,p)=W_{\vartheta,\pm}(r,p)/Z_\pm(\vartheta)\), which are
normalized phase-space densities satisfying
\(\iint P_{\vartheta,\pm}(r,p)\,d^2r\,d^2p=1\).
The negative mass \(Z_-(\vartheta)\) specifies how much negativity is present, while the normalized shape \(P_{\vartheta,-}(r,p)\) specifies where in phase space that negativity is located. In this way the complex entropy naturally separates the total amount of negative weight, encoded in \(h_i(\vartheta)\propto Z_-(\vartheta)\), from the detailed distribution of that weight over phase space, encoded in the negative-channel shape \(P_{\vartheta,-}\).
\begin{figure}
\centering
\includegraphics[width=8.3cm]{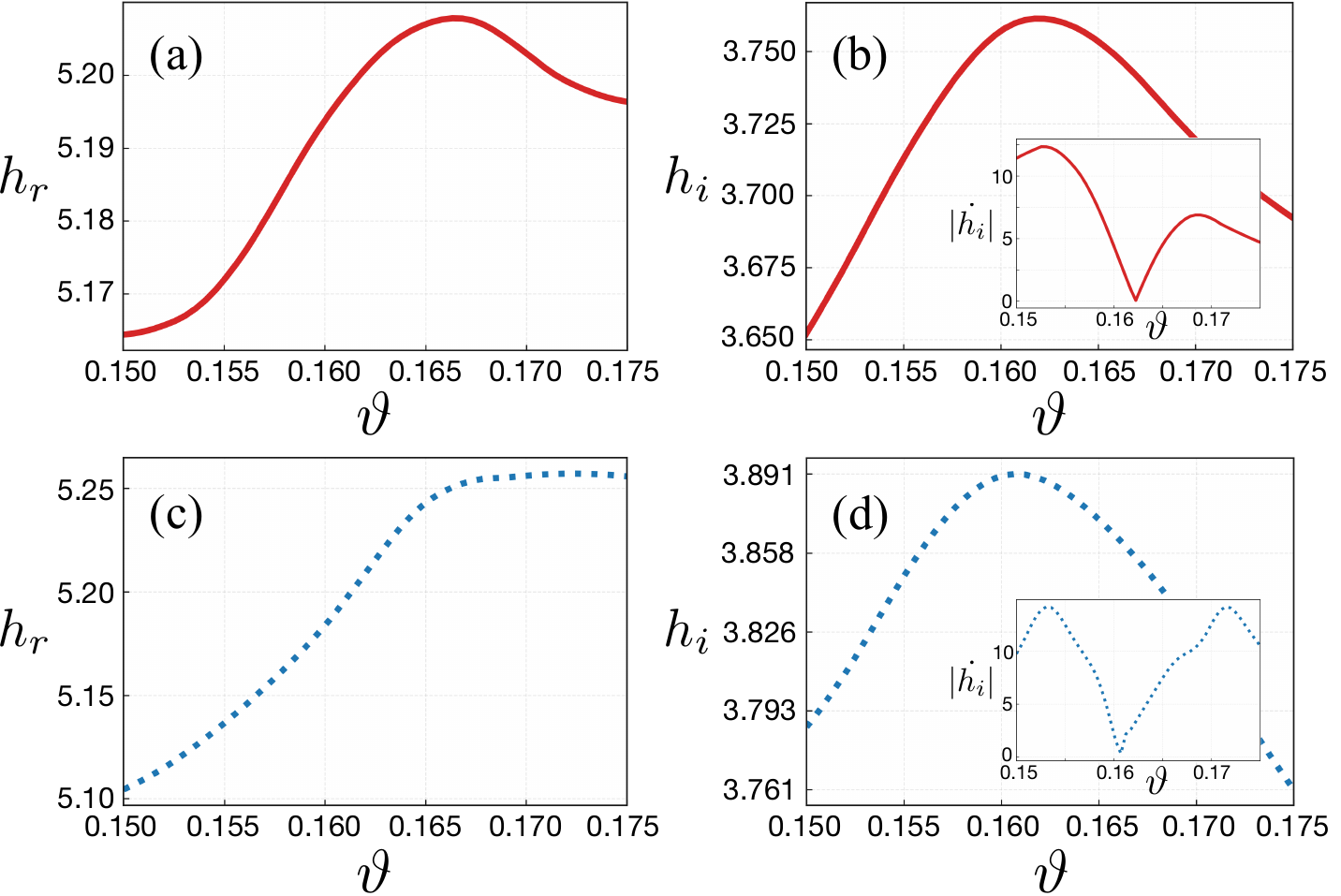}
\caption{
Complex Wigner entropy for the two modes shown in Fig.~1.  
Panels (a) and (b) correspond to mode\,2, displaying the real part \(h_r(\vartheta)\) and the imaginary part \(h_i(\vartheta)\).  
Panels (c) and (d) correspond to mode\,1, showing the same quantities for the lower branch.  
Because \(h_i(\vartheta) \propto N(\vartheta)\) (the Wigner negativity), its local maximum near the center of the avoided crossing reflects the enhancement of phase-space interference.  
The insets in (b) and (d) plot the magnitude \(|dh_i/d\vartheta|\), whose nearly vanishing value marks the precise location of the peak of \(h_i\), allowing clear identification of the center of the avoided crossing (AC).
}
\label{Figure-3}
\end{figure}

These definitions are illustrated in Fig.~\ref{Figure-3} for the two modes participating in the avoided crossing of Fig.~1. For each mode we evaluate the complex entropy \(\mathcal{H}[W_\vartheta]\) along the eigenvalue trajectories and plot its real and imaginary parts. In both modes the imaginary entropy \(h_i(\vartheta)=\pi N(\vartheta)\) exhibits a broad local maximum near the center of the avoided crossing, reflecting the enhancement of phase-space interference and the associated growth of negative Wigner volume. The real part \(h_r(\vartheta)\), by contrast, shows only a gentle variation and does not admit a simple operational interpretation once \(W_\vartheta\) becomes sign-changing. In line with previous work on complex Wigner entropy, this real component is regarded mainly as an auxiliary channel that accompanies \(h_i\), while \(h_i\) carries the physically transparent information about the nonclassicality encoded in Wigner negativity~\cite{CerfComplexWignerEntropy, Kua2025}.

\subsection{Fisher control of negativity and avoided-crossing behavior}

Up to this point the complex Wigner entropy has characterized the static distribution of negativity: \(h_i(\vartheta)\propto Z_-(\vartheta)\) sets the total negative weight, and \(P_{\vartheta,-}(r,p)\) describes how that weight is arranged in phase space. To analyze how this structure reorganizes as the boundary is deformed, we introduce a dynamical measure of sensitivity based on Fisher information. For each sign-resolved channel we define
\begin{equation}
F_\pm(\vartheta)
=\iint d^2r\,d^2p\;
P_{\vartheta,\pm}(r,p)\,
\bigl[\partial_\vartheta\ln P_{\vartheta,\pm}(r,p)\bigr]^2.
\label{eq:Fpm}
\end{equation}
By construction, the quantities \(F_\pm(\vartheta)\) ignore any overall rescaling of \(W_{\vartheta,\pm}\) and respond only to rearrangements of the shapes \(P_{\vartheta,\pm}\). For the negative channel the factorization
\(W_{\vartheta,-}(r,p)=Z_-(\vartheta)\,P_{\vartheta,-}(r,p)\),
together with Eq.~\eqref{eq:neg-def}, shows that the pair \((h_i,F_-)\) captures two complementary aspects of the negativity: \(h_i(\vartheta)=\pi Z_-(\vartheta)\) measures its total amount, whereas \(F_-(\vartheta)\) measures how sensitively the shape of the negative lobe responds to changes in \(\vartheta\).

The construction of shape probabilities \(P_{\vartheta,\pm}\) implies a corresponding decomposition of the local parameter score on the negative lobe: $\partial_\vartheta\ln W_{\vartheta,-}(r,p)
=\partial_\vartheta\ln Z_-(\vartheta)
+\partial_\vartheta\ln P_{\vartheta,-}(r,p).$
If the normalized shape \(P_{\vartheta,-}\) is nearly rigid while only the overall weight \(Z_-(\vartheta)\) changes, then \(\partial_\vartheta\ln P_{\vartheta,-}\approx 0\) and \(F_-(\vartheta)\approx 0\): the amount of negativity can grow without any Fisher response. Conversely, if \(Z_-(\vartheta)\) changes while \(P_{\vartheta,-}\) simultaneously undergoes a rapid rearrangement in phase space, then \(\partial_\vartheta\ln P_{\vartheta,-}\) becomes large and \(F_-(\vartheta)\) develops a pronounced peak. The avoided crossings in Fig.~1 fall into this latter regime.

For later use we introduce a noncentered second moment of the logarithmic amplitude over the negative channel,
\begin{equation}
\widetilde F_-(\vartheta)
:=\mathbb{E}_{P_{\vartheta,-}}\!\bigl[(\partial_\vartheta\ln|W_\vartheta|)^2\bigr],
\label{eq:Ftilde-def}
\end{equation}
where \(\mathbb{E}_{P_{\vartheta,-}}[\cdot]\) denotes expectation with respect to \(P_{\vartheta,-}(r,p)\). On the negative support \(|W_\vartheta|=W_{\vartheta,-}\), and using \(|W_\vartheta|=Z_-(\vartheta)\,P_{\vartheta,-}\) together with the normalization of \(P_{\vartheta,-}\) we obtain
\begin{equation}
\mathbb{E}_{P_{\vartheta,-}}[\partial_\vartheta\ln|W_\vartheta|]
=\partial_\vartheta\ln Z_-(\vartheta)
=\partial_\vartheta\ln N(\vartheta).
\label{eq:score-mean}
\end{equation}
Since
\(
\partial_\vartheta\ln|W_\vartheta|
=\partial_\vartheta\ln Z_-(\vartheta)
+\partial_\vartheta\ln P_{\vartheta,-}(r,p),
\)
adding the constant \(\partial_\vartheta\ln Z_-(\vartheta)\) does not change the variance, so
\(
\mathrm{Var}_{P_{\vartheta,-}}(\partial_\vartheta\ln|W_\vartheta|)
=\mathrm{Var}_{P_{\vartheta,-}}(\partial_\vartheta\ln P_{\vartheta,-})
=F_-(\vartheta).
\)
Combining Eqs.~\eqref{eq:Fpm} and \eqref{eq:score-mean} gives the variance decomposition
\begin{align}
\widetilde F_-(\vartheta)
&=\mathrm{Var}_{P_{\vartheta,-}}\!\bigl(\partial_\vartheta\ln|W_\vartheta|\bigr)
+\bigl[\partial_\vartheta\ln N(\vartheta)\bigr]^2
\notag\\[2pt]
&=F_-(\vartheta)
+\left[\frac{N'(\vartheta)}{N(\vartheta)}\right]^{\!2},
\label{eq:Ftilde-decomp}
\end{align}
where
\(
F_-(\vartheta)
=\mathrm{Var}_{P_{\vartheta,-}}(\partial_\vartheta\ln|W_\vartheta|)
\)
is the centered Fisher information of the negative channel and the additional term captures the trivial growth rate of the total negative mass.

We now derive a Fisher-type control on the slope of the imaginary entropy \(h_i(\vartheta)=\pi N(\vartheta)\). Differentiating the negativity volume and rewriting the result as a \(P_{\vartheta,-}\)-expectation gives
\begin{equation}
\frac{dN}{d\vartheta}
=\iint_{W_\vartheta<0}\partial_\vartheta|W_\vartheta|\,d^2r\,d^2p
=N(\vartheta)\,\mathbb{E}_{P_{\vartheta,-}}\!\bigl[\partial_\vartheta\ln|W_\vartheta|\bigr].
\label{eq:dN}
\end{equation}
Applying the Cauchy--Schwarz inequality in the probability space defined by \(P_{\vartheta,-}\) to the random variable \(X(r,p)=\partial_\vartheta\ln|W_\vartheta(r,p)|\) yields
\(|\mathbb{E}_{P_{\vartheta,-}}[X]|\le\sqrt{\mathbb{E}_{P_{\vartheta,-}}[X^2]}\), which directly leads to the Fisher control 
\begin{equation}
\bigg|\frac{dh_i}{d\vartheta}\bigg|
\le
\pi\,N(\vartheta)\,\sqrt{\widetilde F_-(\vartheta)}.
\label{eq:hi-control}
\end{equation}
This inequality shows that the slope of the imaginary entropy is constrained by two distinct ingredients: the total negative mass \(N(\vartheta)\), which sets the overall scale, and the noncentered Fisher quantity \(\widetilde F_-(\vartheta)\), which measures the magnitude of local score fluctuations \(\partial_\vartheta\ln|W_\vartheta|\) over the negative lobe.

From Eqs.~\eqref{eq:Ftilde-decomp} and \eqref{eq:dN} we can interpret these quantities more transparently. On the negative lobe the score field
\(
S_\vartheta(r,p) = \partial_\vartheta\ln|W_\vartheta(r,p)|
\)
has mean
\(
\mathbb{E}_{P_{\vartheta,-}}[S_\vartheta]
= \partial_\vartheta\ln N(\vartheta),
\)
that is, the logarithmic growth rate of the total negativity. Its variance is
\(
F_-(\vartheta)
=\mathrm{Var}_{P_{\vartheta,-}}(S_\vartheta),
\)
which coincides with the negative-channel Fisher information and measures the strength of score fluctuations over the negative support. The decomposition \(\widetilde F_-(\vartheta)=F_-(\vartheta)+[\partial_\vartheta\ln N(\vartheta)]^2\) then makes explicit that \(\widetilde F_-(\vartheta)\) combines a contribution from the global growth of the negative mass with a genuinely shape-sensitive contribution from spatial variations of the score field.

At the center \(\vartheta_0\) of an avoided crossing, the numerical data in Fig.~\ref{Figure-3} show that \(h_i(\vartheta)\) has a smooth local maximum, so that \(\left.\partial_\vartheta\ln N(\vartheta)\right|_{\vartheta_0}=0\). In this case the noncentered and centered Fisher quantities coincide, \(\widetilde F_-(\vartheta_0)=F_-(\vartheta_0)\), and the entire score field \(\partial_\vartheta\ln|W_\vartheta|\) is generated by shape changes of the negative lobe rather than by changes of its total mass.

\begin{figure}
\centering
\includegraphics[width=8.2cm]{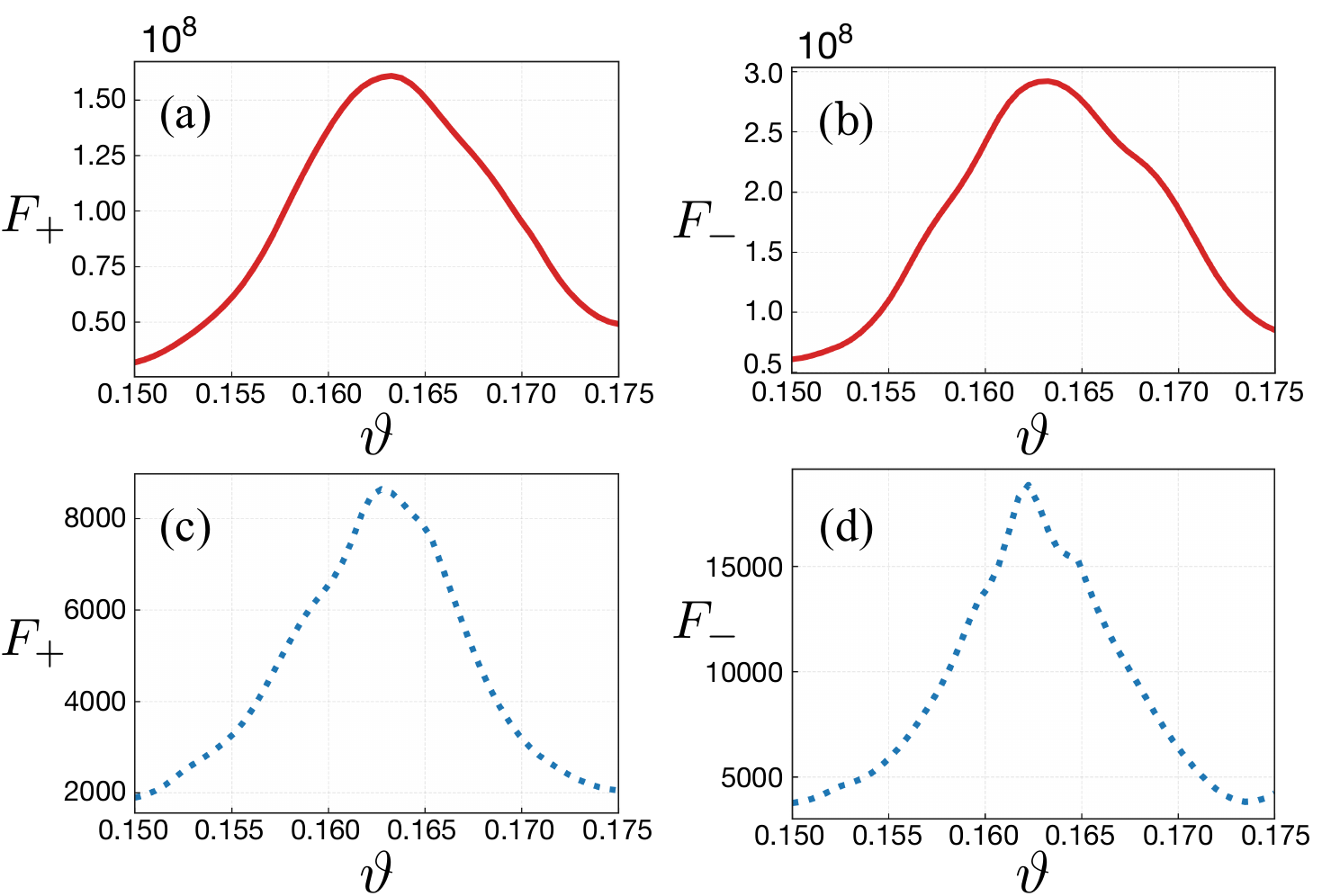}
\caption{
Positive- and negative-channel Fisher informations for the two modes in Fig.~1.  
Panels (a) and (b) correspond to mode\,2, showing the positive-channel \(F_{+}(\vartheta)\) and negative-channel \(F_{-}(\vartheta)\) on a vertical scale of \(10^{8}\).  
Panels (c) and (d) display the same quantities for mode\,1.  
In both modes \(F_{-}(\vartheta)\) is systematically larger, typically exceeding \(F_{+}(\vartheta)\) by more than a factor of two, which indicates that the negative Wigner lobe undergoes a stronger parameter-induced reshaping.  
Both \(F_{+}(\vartheta)\) and \(F_{-}(\vartheta)\) exhibit clear peaks near the avoided-crossing region, providing a sharp phase-space signature of mode hybridization that is consistent with the maxima of \(h_i(\vartheta)\) in Fig.~\ref{Figure-3}.
}
\label{Figure-4}
\end{figure}

Taken together, Eqs.~\eqref{eq:complex-entropy-def}--\eqref{eq:hi-control} and Figs.~\ref{Figure-3} and \ref{Figure-4} provide a channel-resolved description in which the imaginary entropy \(h_i(\vartheta)=\pi N(\vartheta)\) quantifies how much nonclassicality is present, while the negative-channel Fisher information \(F_-(\vartheta)\) quantifies how rapidly the shape of the negative lobe rearranges as \(\vartheta\) is varied. The Fisher control~\eqref{eq:hi-control} then links these two aspects by stating that the rate of change of \(h_i(\vartheta)\) cannot exceed an envelope set by the product of the total negative mass and the root-mean-square score \(\sqrt{\widetilde F_-(\vartheta)}\).

It is useful to contrast the Fisher control~\eqref{eq:hi-control} with the standard Cramér--Rao inequality. If the normalized negative-channel shapes \(P_{\vartheta,-}\) are regarded as a statistical model for measurement outcomes drawn from the negative lobe, then the Fisher information \(F_-(\vartheta)\) enters the Cramér--Rao bound for any unbiased estimator of a scalar function \(g(\vartheta)\), providing a lower limit of the form \(\mathrm{Var}[T]\ge[g'(\vartheta)]^{2}/F_-(\vartheta)\) for the variance of a single-sample estimator~\cite{{CoverThomas}}. Equation~\eqref{eq:hi-control}, by contrast, is a deterministic control on the slope of \(h_i(\vartheta)\) obtained from a Cauchy--Schwarz inequality in the space of phase-space scores. In other words, \(F_-(\vartheta)\) plays a dual role: in the statistical interpretation it bounds the best possible precision with which functions of \(\vartheta\) such as \(N(\vartheta)\) or \(h_i(\vartheta)\) could in principle be inferred from negative-lobe data, while in the present phase-space analysis it enters \(\widetilde F_-(\vartheta)\) and combines with \(N(\vartheta)\) to constrain how sharply the imaginary entropy itself can vary across an avoided crossing.

\section{Conclusion}

We have developed a channel-resolved, information-theoretic description of Wigner negativity based on the complex Gibbs--Shannon functional and applied it to avoided crossings in an oval quantum billiard. For a real, sign-changing Wigner function this functional becomes complex, and its imaginary part, proportional to the Kenfack--\.Zyczkowski negative volume, provides an entropy-like measure of phase-space nonclassicality. A sign-resolved decomposition into positive and negative channels separates the total amount of negativity from its normalized shape and identifies where in phase space the negative weight concentrates as the control parameter \(\vartheta\) is varied.

On this basis we introduced Fisher information for the channel shapes, with the negative-channel quantity \(F_-(\vartheta)\) quantifying how sensitively the negative Wigner lobe reshapes under boundary deformations. A Cauchy--Schwarz argument yields a Fisher-type bound that limits how rapidly the imaginary entropy, and hence the Wigner negativity, can change with \(\vartheta\), tying its slope to the product of the total negative mass and the score fluctuations on the negative support. In the avoided crossing studied here this framework shows that mode hybridization is accompanied by a broad peak of \(h_i(\vartheta)\) and a sharp enhancement of \(F_-(\vartheta)\), providing a phase-space signature of the mixing. Although our numerical illustration is a single oval billiard, the construction is generic for systems with phase-space representations and offers a natural starting point for extensions to more complex wave-chaotic, mesoscopic, or non-Hermitian settings.

\section*{Acknowledgment}
This work was supported by the National Research Foundation of Korea (NRF) through a grant funded by the Ministry of Science and ICT (Grant Nos. RS-2023-00211817, RS-2025-00515537, and RS-2022-NR068791).

\section*{Data availability}
The data are not publicly available. The data are available from the authors upon reasonable request.

\section*{Declaration of competing interest}
The authors declare no conflicts of interest

\end{document}